\documentclass[copyright,creativecommons]{eptcs}
\providecommand{\event}{GandALF 2011} 
\usepackage{breakurl}
\usepackage{amssymb,amsmath,amscd,latexsym}
\usepackage[vlined,boxruled]{algorithm2e}
\usepackage{pstricks}
\usepackage{macro}
\usepackage{graphicx}
\usepackage{subfigure}

\newtheorem{theorem}{Theorem}

\title{Towards Efficient Exact Synthesis for Linear Hybrid Systems}
\author{Massimo Benerecetti \and Marco Faella
\institute{Universit\`a di Napoli\\``Federico II'', Italy}
\email{\{mfaella,bene,minopoli\}@na.infn.it}
\and Stefano Minopoli
}
\def\titlerunning{Towards Efficient Exact Synthesis for Linear Hybrid Systems}
\def\authorrunning{M. Benerecetti, M. Faella \& S. Minopoli}
\begin{document}
\maketitle

\begin{abstract}
We study the problem of automatically computing the controllable
region of a Linear Hybrid Automaton, with respect to a safety objective. 
We describe the techniques that are needed to effectively
and efficiently implement a recently-proposed solution procedure,
based on polyhedral abstractions of the state space.
Supporting experimental results are presented, based on an
implementation of the proposed techniques on top of the tool PHAVer.
\end{abstract}

\section{Introduction}

Hybrid systems are an established formalism for modeling
physical systems which interact with a digital controller.
From an abstract point of view, a hybrid system is a dynamic system
whose state variables are both discrete and continuous.
Typically, continuous variables represent physical quantities like
temperature, speed, etc., while discrete ones represent \emph{control modes},
i.e., states of the controller.

Hybrid automata~\cite{hybridLics96} are the most common syntactic variety of hybrid system:
a finite set of locations, similar to the states of a finite automaton,
represents the value of the discrete variables.
The current location, together with the current value of 
the (continuous) variables, form the instantaneous description of the system.
Change of location happens via discrete transitions, and 
the evolution of the variables is governed 
by differential equations attached to each location.
In a Linear Hybrid Automaton (LHA), the allowed differential
equations are in fact \emph{polyhedral differential inclusions}
of the type $\dot{\mathbf{x}} \in P$, where $\dot{\mathbf{x}}$ 
is the vector of the first derivatives of all variables
and $P$ is a convex polyhedron.
Notice that differential inclusions are non-deterministic, allowing
for infinitely many solutions.

We study LHAs whose discrete transitions are partitioned into
controllable and uncontrollable ones, and we wish to compute a
strategy for the controller to satisfy a given goal, regardless of the
evolution of the continuous variables and of the uncontrollable
transitions.  Hence, the problem can be viewed as a \emph{two player
  game}: on one side the controller, who can only issue controllable
transitions, on the other side the environment, who can choose the
trajectory of the variables and can take uncontrollable transitions at
any moment.

As control goal, we consider safety, i.e., the objective of keeping
the system within a given region of safe states.  This problem has
been considered several times in the literature.
In~\cite{lha_techrep11},
%\footnote{The technical report~\cite{lha_techrep11}, 
%  which is strictly connected with the present paper, 
%  is currently submitted to a major
%  conference.},
we fixed some inaccuracies in previous presentations,
and proposed a sound and complete semi-procedure for the problem.
Here, we discuss the techniques required to efficiently implement the
algorithms in~\cite{lha_techrep11}.  In particular, two operators on
polyhedra need non-trivial new developments to be exactly and efficiently
computed.  Both operators pertain to intra-location behavior,
and therefore assume that trajectories are subject to a fixed
polyhedral differential inclusion of the type $\dot{\mathbf{x}} \in
P$.
\begin{itemize}
\item The \emph{pre-flow} operator. Given a polyhedron
  $U\subseteq\reals^n$, we wish to compute the set of all points that
  may reach $U$ via an admissible trajectory. This apparently easy
  task becomes non-trivial when the convex polyhedron $P$ is not
  (necessarily) topologically closed.  This is the topic of
  Section~\ref{sec:preflow}.
\item The \emph{may reach while avoiding} operator, denoted by $\rwamay$.
  Given two polyhedra $U$ and $V$, the operator computes the set of
  points that may reach $U$ while avoiding $V$, via an admissible
  trajectory.  A fixpoint algorithm for this operator was presented
  in~\cite{lha_techrep11}.  Here, we introduce a number of efficiency
  improvements (Section~\ref{sec:sorm}), accompanied by a
  corresponding experimental evaluation
  (Section~\ref{sec:experiments}), carried out on our tool PHAVer+,
  based on the open-source tool PHAVer~\cite{phaver05}.
\end{itemize}
Contrary to most recent literature on the subject, we focus on exact 
algorithms. Although it is established that exact analysis and synthesis
of realistic hybrid systems is computationally demanding, %prohibitive?
we believe that the ongoing research effort on approximate techniques
should be based on the solid grounds provided by the exact approach.
For instance, a tool implementing an exact algorithm (like our PHAVer+)
may serve as a benchmark to evaluate the performance and the precision
of an approximate tool.

\paragraph{Related work.}
The idea of automatically synthesizing controllers for dynamic systems
first arose in connection with discrete systems~\cite{RW87}.
%similar to finite automata~\cite{RW87}.
Then, the same idea was applied to real-time systems modeled by timed
automata~\cite{MalerGames}, thus coming one step closer to the
continuous systems that control theory usually deals with.  Finally,
it was the turn of hybrid systems~\cite{WongToi,hybridgames99}, and in
particular of LHA, the very model that we analyze
in this paper. Wong-Toi proposed the first symbolic semi-procedure to
compute the controllable region of a LHA w.r.t. a safety
goal~\cite{WongToi}.  The heart of the procedure lies in the operator
\emph{flow\_avoid}$(U,V)$, which is analogous to our $\rwamay$.
However, the algorithm provided in~\cite{WongToi} 
for \emph{flow\_avoid} does not work for non-convex
$V$, a case which is very likely to occur in practice, even if the
original safety goal is convex. A revised algorithm, correcting such flaw, was 
proposed in~\cite{lha_techrep11}. 

Tomlin et al.\ and Balluchi et al.\ analyze much more expressive
models~\cite{tomlin00,Balluchi03}, with generality in mind rather than
automatic synthesis. Their \emph{Reach} and
\emph{Unavoid\_Pre} operators, respectively, again correspond to
$\rwamay$.

Asarin et al. investigate the synthesis problem for hybrid systems
where all discrete transitions are controllable and the trajectories
satisfy given linear differential equations of the type
$\dot{\mathbf{x}} = A \mathbf{x}$~\cite{asarin00}.  The expressive
power of these constraints is incomparable with the one offered by the
differential inclusions occurring in LHAs.  In particular, linear
differential equations give
rise to deterministic trajectories, while differential inclusions are
non-deterministic.  In control theory terms, differential inclusions
can represent the presence of environmental \emph{disturbances}. The
tool {\tt d/dt}~\cite{ddt02}, by the same authors, is reported to
support controller synthesis for safety objectives, but the publicly
available version in fact does not.

\begin{comment}
The rest of the paper is organized as follows.  Section~\ref{sec:defs}
introduces and motivates the model.  In Section~\ref{sec:algo}, we
present the semi-procedure which solves the synthesis problem.
Section~\ref{sec:preflow} presents an algorithm for computing the
pre-flow of a given polyhedron. Section~\ref{sec:sorm}
describes...
the reports some experiments performed on
Section~\ref{sec:experiments} reports some experiments performed on
our implementation of the procedure and Section~\ref{sec:conclusions}
draws some conclusions.
% Concluding remarks and future
%work are given in Section~\ref{sec:conclusion}.
\end{comment}

\section{Linear Hybrid Automata} \label{sec:defs}

A \emph{convex polyhedron} is a subset of $\reals^n$ that is the intersection
of a finite number of half-spaces.
A \emph{polyhedron} is a subset of $\reals^n$ that is 
the union of a finite number of convex polyhedra.
For a general (i.e., not necessarily convex) polyhedron $G \subseteq \reals^n$, 
we denote by $\conv{G} \subseteq 2^{\reals^n}$ the finite set of convex polyhedra comprising it.
%
%In a \emph{Linear Hybrid Automaton (LHA)}, invariants
%and initial states are given by polyhedra over $X$, flow predicates by convex
%polyhedra over $\dot{X}$, and jump relations by polyhedra over $X \cup X'$.

Given an ordered set $X = \{x_1, \ldots , x_n\}$ of variables, a \emph{valuation} 
is a function $v : X\rightarrow\reals$.
Let $\Val(X)$ denote the set of valuations over $X$.
There is an obvious bijection between $\Val(X)$ and $\reals^n$,
allowing us to extend the notion of (convex) polyhedron to sets of valuations.
We denote by $\cpoly(X)$ (resp., $\poly(X)$) the set of convex polyhedra 
(resp., polyhedra) on $X$.

%When a valuation $u$ over $X$ and a valuation $v$ over $Y$ agree on the shared
%variables, i.e., $u(x) = v(x)$ for all $x \in X \cap Y$, 
%we use $u\sqcup v$ to denote the \emph{joint} valuation $w$ over $X \cup Y$
%defined by $w(x) = u(x)$ if $x \in X$ and $w(x) = v(x)$ otherwise.
We use $\dot{X}$ to denote the set $\{\dot{x}_1, \ldots , \dot{x}_n\}$
of dotted variables, used to represent the first derivatives, and $X'$
to denote the set $\{x'_1, \ldots , x'_n\}$ of primed variables, used
to represent the new values of variables after a transition.
Arithmetic operations on valuations are defined in the straightforward way.
An \emph{activity} over $X$ is a differentiable function 
$f:\reals^{\geq 0}\rightarrow \Val(X)$.
%such that its first derivative is Lipschitz continuous.
% The role of the Lipschitz continuity assumption
%will become clear in Section~\ref{sec:reach}.
%\mynote{Serve Lipschitz continuity?}
Let $Acts(X)$ denote the set of activities over $X$.
The \emph{derivative} $\dot{f}$ of an activity $f$ is defined in the standard way and
it is an activity over $\dot{X}$.
%defined analogously to the derivative in $\reals^n$.
%The extension of operators from valuations to activities is done pointwise.
%Let $const_X(Y) = \{(v, v') \vert v, v'\in V (X), v\proj_Y = v'\proj_Y\}$.
A \emph{Linear Hybrid Automaton} $H = (\Loc, X, \edgc, \edgu, \Flow,
\Inv, \Init)$ consists of the following:
\begin{itemize}
\item A finite set $\Loc$ of \emph{locations}.
\item A finite set $X = \{x_1, \ldots, x_n\}$ of continuous,
  real-valued \emph{variables}.  A \emph{state} is a pair $(l, v)$ of
  a location $l$ and a valuation $v \in \Val(X)$.
\item Two sets $\edgc$ and $\edgu$ of \emph{controllable} and
  \emph{uncontrollable transitions}, respectively. They describe
  instantaneous changes of locations, in the course of which variables
  may change their value.  Each transition $(l, \mu, l')\in \edgc \cup
  \edgu$ consists of a \emph{source location} $l$, a \emph{target
    location} $l'$, and a \emph{jump relation} $\mu \in \poly(X \cup
  X')$, that specifies how the variables may change their value during
  the transition.  The projection of $\mu$ on $X$ describes the
  valuations for which the transition is enabled; this is often
  referred to as a \emph{guard}.
\item A mapping $\Flow : \Loc \to \cpoly(\dot{X})$ attributes to each
  location a set of valuations over the first derivatives of the
  variables, which determines how variables can change over time.
\item A mapping $\Inv : \Loc \to \poly(X)$, called the
  \emph{invariant}.
\item A mapping $\Init: \Loc \to \poly(X)$, contained in the
  invariant,, which allows the definition of the \emph{initial states}
  from which all behaviors of the automaton originate.
\end{itemize}
We use the abbreviations $S = \Loc \times \Val(X)$ for the set of states
and $\edg = \edgc \cup \edgu$ for the set of all transitions.
Moreover, we let $\Invs = \bigcup_{l\in\Loc} \{ l \} \times \Inv(l)$
and $\Inits = \bigcup_{l\in\Loc} \{ l \} \times \Init(l)$.
Notice that $\Invs$ and $\Inits$ are sets of states.

\subsection{Semantics}

The behavior of a LHA is based on two types
of transitions: \emph{discrete} transitions correspond
to the $\edg$ component, and produce an instantaneous change
in both the location and the variable valuation;
\emph{timed} transitions describe the change of the variables over
time in accordance with the $\Flow$ component.

Given a state $s = \pair{l, v}$, we set $\loc(s) = l$ and $\val(s) = v$.
An activity $f \in Acts(X)$ is called \emph{admissible from $s$}
if \emph{(i)} $f(0) = v$ and
\emph{(ii)} for all $\delta\geq 0$ it holds $\dot{f}(\delta) \in \Flow(l)$.
We denote by $\adm(s)$ the set of activities that are admissible from $s$.
Additionally, for $f \in \adm(s)$, the \emph{span} of $f$ in $l$, denoted by $\sp(f,l)$
is the set of all values $\delta\geq 0$
such that $\pair{l, f(\delta')} \in \Invs$ for all $0\leq\delta'\leq\delta$.
Intuitively, $\delta$ is in the span of $f$ iff $f$ never leaves the invariant
in the first $\delta$ time units.
If all non-negative reals belong to $\sp(f,l)$, we write $\infty \in \sp(f,l)$.

\paragraph{Runs.}
Given two states $s, s'$, and a transition $e \in \edg$,
there is a \emph{discrete transition} $s \xto{e} s'$ 
with \emph{source} $s$ and \emph{target} $s'$ iff
\emph{(i)} $s, s' \in Invs$, 
\emph{(ii)} $e = (loc(s), \mu, loc(s'))$, and
\emph{(iii)} $(val(s), val(s')')\in \mu$, where $\val(s')'$ is the valuation
over $X'$ obtained from $\val(s')$ by renaming each variable $x \in X$ onto the corresponding primed variable $x' \in X$.
There is a \emph{timed transition} $s \xto{\delta,f} s'$ with \emph{duration} 
$\delta\in\reals^{\geq 0}$ and activity $f \in \adm(s)$ iff 
\emph{(i)} $s \in Invs$, 
\emph{(ii)} $\delta \in \sp(f,\loc(s))$, and
\emph{(iii)} $s' = \pair{\loc(s), f(\delta)}$.
For technical convenience, we admit timed transitions of duration zero%
\footnote{Timed transitions of duration zero can be disabled by
adding a clock variable $t$ to the automaton and requesting that each
discrete transition happens when $t>0$ and resets $t$ to $0$ when taken.}.
A special timed transition is denoted $s \xto{\infty,f}$
and represents the case when the system follows an activity forever.
This is only allowed if $\infty \in \sp(f,\loc(s))$.
Finally, a \emph{joint transition} $s \xto{\delta,f,e} s'$ 
represents the timed transition $s \xto{\delta,f} \pair{\loc(s),f(\delta)}$
followed by the discrete transition $\pair{\loc(s),f(\delta)} \xto{e} s'$.

A \emph{run} is a sequence
\begin{equation} \label{eq:run} 
r = s_0 \xto{\delta_0,f_0} s_0' \xto{e_0}
      s_1 \xto{\delta_1,f_1} s_1' \xto{e_1} s_2 \ldots s_n \ldots
%$r = s_0\xto{\delta_0,f_0,a_0} s_1
%        \xto{\delta_1,f_1,a_1} \ldots \xto{\delta_{n-1},f_{n-1},a_{n-1}} s_n \ldots$
\end{equation}
of alternating timed and discrete transitions, such that either the
sequence is infinite, or it ends with a timed transition of the type
$s_n \xto{\infty,f}$.  If the run $r$ is finite, we define
$\length(r)=n$ to be the length of the run, otherwise we set
$\length(r)=\infty$.  The above run is \emph{non-Zeno} if for all
$\delta\geq 0$ there exists $i\geq 0$ such that $\sum_{j=0}^{i}
\delta_j > \delta$.  We denote by $\states(r)$ the set of all states
visited by $r$.  Formally, $\states(r)$ is the
set of states $\pair{\loc(s_{i}), f_{i}(\delta)}$, for all $0
\leq i \leq \length(r)$ and all $0 \leq \delta \leq \delta_i$.
%
%For a non-Zeno run $r$ and a time $\delta \geq 0$, with an abuse of notation we denote
%by $r(\delta)$ the state encountered by the run at time $\delta$.
%Formally, let $i$ be such that 
%$\sum_{j=0}^{i-1} \delta_j \leq \delta < \sum_{j=0}^{i} \delta_j$,
%we have $r(\delta) = \pair{\loc(s_{i}), f_{i}(\delta - \sum_{j=0}^{i-1} \delta_j)}$. The function $r: \reals^{\geq 0} \to S$ so defined can be called
%the \emph{flat view} of the run $r$.
%
Notice that the states from which discrete transitions start
(states $s_i'$ in~\eqref{eq:run}) appear in $\states(r)$.
Moreover, if $r$ contains a sequence of one or more
zero-time timed transitions, all intervening states appear in $\states(r)$.

\paragraph{Zenoness and well-formedness.}
A well-known problem of real-time and hybrid systems is that
definitions like the above admit runs that take infinitely many
discrete transitions in a finite amount of time (i.e., \emph{Zeno}
runs), even if such behaviors are physically meaningless.  In this
paper, we assume that the hybrid automaton under consideration
generates no such runs. This is easily achieved by using an extra
variable, representing a clock, to ensure that the delay between any
two transitions is bounded from below by a constant.
%adding an extra variable $t$,
%representing a clock, to the model, and ensuring that all transitions
%can only be taken when $t\geq 1$ and that they reset $t$ to zero.
We leave it to future work to combine our results with more
sophisticated approaches to Zenoness known in the
literature~\cite{Balluchi03,dAFHMS03}.

Moreover, we assume that the hybrid automaton under consideration
is \emph{non-blocking}, i.e., whenever the automaton is about
to leave the invariant there must be an uncontrollable transition
enabled.
%Formally, for all states $s$ in the invariant,
%if all activities $f \in \adm(s)$ eventually leave the invariant,
%there exists one such activity $f$ and a time $\delta \in \sp(f, \loc(s))$
%such that $s' = \pair{\loc(s), f(\delta)}$ 
%is in the invariant and there is an uncontrollable transition 
%$e \in \edgu$ such that $s' \xto{e} s''$.
If a hybrid automaton is non-Zeno and non-blocking, we say that
it is \emph{well-formed}.
In the following, all hybrid automata are assumed to be well-formed.

\paragraph{Strategies.}
%\mynote{Why not having strategies choose either an action or a delay?
%Check ``The element of surprise''.}
%
A \emph{strategy} is a function
$\sigma: S \to 2^{\edgc \cup \{\bot\}} \setminus \emptyset$,
where $\bot$ denotes the null action.
Notice that our strategies are \emph{non-deterministic} and \emph{memoryless}
(or \emph{positional}).
A strategy can only choose a transition which is allowed by the automaton.
Formally, for all $s\in S$, if $e \in \sigma(s) \cap \edgc$, 
then there exists $s'\in S$ such that $s\xto{e} s'$.
Moreover, when the strategy chooses the null action,
it should continue to do so for a positive amount of time,
along each activity that remains in the invariant.
If all activities immediately exit the invariant,
the above condition is vacuously satisfied.
%
%Formally, if $\bot \in \sigma(s)$,
%for all $f \in \adm(s)$ there exists $\delta>0$ such that
%for all $0 < \delta' < \delta$ it holds 
%$\delta' \not\in \sp(f, \loc(s))$ or  
%$\bot \in \sigma(\pair{\loc(s),f(\delta')})$.
This ensures that the null action
is enabled in right-open regions, so that there is an earliest instant
in which a controllable transition becomes mandatory. 

Notice that a strategy can always choose the null action.
The well-formedness condition ensures that the system can always
evolve in some way, be it a timed step or an uncontrollable transition.
In particular, even if we are on the boundary of the invariant
we allow the controller to choose the null action, because,
in our interpretation,
it is not the responsibility of the controller
to ensure that the invariant is not violated.

We say that a run like~\eqref{eq:run} is \emph{consistent} with a strategy 
$\sigma$ if for all $0\leq i< \length(r)$ the following conditions hold:
\begin{itemize}
\item for all $\delta \geq 0$ such that
$\sum_{j=0}^{i-1} \delta_j \leq \delta < \sum_{j=0}^{i} \delta_j$,
we have $\bot \in \sigma(\pair{\loc(s_i), f_i(\delta - \sum_{j=0}^{i-1} \delta_j)})$;
\item if $e_i \in \edgc$ then $e_i \in \sigma(s_i')$.
\end{itemize}
We denote by $\runs(s, \sigma)$ the set of runs starting from the state $s$
and consistent with the strategy $\sigma$.

\paragraph{Safety control problem.}
Given a hybrid automaton and a set of states $T\subseteq\Invs$,
the \emph{safety control problem} asks whether
there exists a strategy $\sigma$ such that, for all initial states $s \in \Inits$,
all runs $r \in \runs(s, \sigma)$
it holds $\states(r) \subseteq T$.
%We call the above $\sigma$ a \emph{winning strategy}.

\section{Solving the Safety Control Problem} \label{sec:algo}

In this section, we recall the semi-procedure that solves the safety
control problem for a given LHA and safe region.
It is well known in the literature (see e.g.~\cite{maler02,asarin00}) that
the answer to the safety control problem for safe
set $T\subseteq\Inv$ is positive if and only if
$$\Init \subseteq \nu W \qdot T \cap \cpre(W),$$
where $\cpre$ is the {\it controllable predecessor operator}, defined
below.
Since the reachability problem for LHA was proved undecidable~\cite{whatsdecidable98},
the above fixpoint may not converge in a finite number of steps.
On the other hand, it does converge in many cases of practical interest,
as witnessed by the examples in Section~\ref{sec:experiments}.

For a set of states $A$, the operator $\cpre(A)$ returns the set of
states from which the controller can ensure that the system remains in
$A$ during the next joint transition.  This happens if for all
activities chosen by the environment and all delays $\delta$, one of
two situations occurs:
\begin{itemize}
\item either the systems stays in $A$ up to time $\delta$, while all
  uncontrollable transitions enabled up to time $\delta$ (included)
  also lead to $A$, or
\item some preceding instant $\delta' < \delta$ exists such that the
  system stays in $A$ up to time $\delta'$, while all uncontrollable
  transitions enabled up to time $\delta'$ (included) also lead to
  $A$, and the controller can issue a transition at time $\delta'$
  leading to $A$.
\end{itemize}

%This may happen due to two different situations.  First, there are
%states from which, depending on the activity chosen by the
%environment, one of two cases occurs.  For some activities, the system
%will stay in $A$ forever, even if the controller does not act.  For
%the others, after some delay $\delta$ either there must be some
%controllable transition enabled leading to $A$ or a state in $\urgent$
%is reached, where the only option for the environment is to take a
%transition leading into $A$; moreover, the system remains in $A$
%before $\delta$, and all uncontrollable transitions enabled up to time
%$\delta$ (included) also lead to $A$.

In order to compute $\cpre(A)$ on LHA, the auxiliary operator $\rwamay$
(\emph{may reach while avoiding}) was proposed~\cite{lha_techrep11}.
Intuitively, 
given a location $l$ and two sets of variable valuations $U$ and $V$,
$\rwamay_l(U,V)$ contains the set of valuations from which the
continuous evolution of the system \emph{may} reach $U$ while avoiding
$V \cap \overline{U}$.
%~\footnote{In $\atl$ notation, we have
%  $\rwamay(U,V) \equiv \team{env} (\overline{V} \cup U) \U U$, where
%  $env$ is the player representing the environment.}.
%
% Notice that on a dense time domain this is not equivalent to
% reaching $U$ while avoiding $V$: If an activity avoids $V$ in a
% right-closed interval, and then enters $U \cap V$, the first
% property holds, while the latter does not.

For a set of states $A$ and $x \in \{ u, c\}$, let $\pre{x}{m}(A)$
(for \emph{may predecessors}) be the set of states where some discrete
transition leading to $A$ and belonging to $\edg_x$ is enabled. We
denote with $\proj{A}{l}$ the projection of $A$ on $l$, i.e. $\{v\in
\Val(X) \mid \pair{l,v}\in A\}$.
As proved in~\cite{lha_techrep11}, we then have that 
%
%\begin{multline*}
$$\cpre(A) = \bigcup_{l \in \Loc} \{ l \} \times \Big( \proj{A}{l} \setminus
\rwamay_l\big(\Inv(l) \cap \big( \overline{\proj{A}{l}} \cup B_l \big),
               C_l \cup \overline{\Inv(l)} \big) \Big),$$
%\end{multline*}
%\cpre(A) = \bigcup_{l \in
%  \Loc} \{ l \} \times \Big( \proj{A}{l} \setminus \rwamay_l\big(
%\overline{\proj{A}{l}} \cup B_l, C_l \cup D_l \big) \Big).$$
where $B_l = \proj{\pre{u}{m}\big(\overline{A}\big)}{l}$ and $C_l =
\proj{\pre{c}{m}(A)}{l}$.

Intuitively, the set $B_l$ is the set of valuations $u$ such that from
state $\pair{l, u}$ the environment can take a discrete transition
leading outside $A$, and $C_l$ is the set of valuations $u$ such that
from $\pair{l, u}$ the controller can take a discrete transition into
$A$.  Then, using the $\rwamay$ operator, we compute the set of
valuations from which there exists an activity that either leaves $A$
or enters $B_l$, while staying in the invariant and avoiding $C_l$.
These valuations do not belong to $\cpre(A)$, as the environment can
violate the safety goal within (at most) one discrete transition.

Next, we show how to characterize $\rwamay$ in terms of simple
operations on polyhedra.
Let $\clos{P}$ denote the topological closure of a polyhedron $P$.
Given two polyhedra $P$ and $F$, the \emph{pre-flow} of $P$
w.r.t.\ $F$ is:
$$\pref{P}{F} = \{ x - \delta y \mid x \in P, y \in F, \delta\geq 0 \}.$$
For a given location $l\in\Loc$, the pre-flow of $P$ w.r.t.\ $\Flow(l)$
is the set of points that can reach $P$
via a straight-line activity whose slope is allowed in $l$.
For notational convenience, we use the abbreviation $\lpref{P}$
for $\pref{P}{\Flow(l)}$, and for all polyhedra $P$ and $P'$ we
define their \emph{boundary} to be
$$\bound(P, P') = (\clos{P} \cap P') \cup (P \cap \clos{P'}),$$ 
which identifies a boundary between two (not necessarily closed)
convex polyhedra. Clearly, $\bound(P, P')$ is not empty only if $P$ and
$P'$ are adjacent to one another or if they overlap; it is empty,
otherwise. Moreover, given a location $l$, $entry(P,P')$, the {\it
  entry region} between $P$ and $P'$, denotes the set of points of the
boundary between $P$ and $P'$ which can reach $P'$ by following some
straight-line activity in location $l$. In symbols:
$entry(P,P') =  \bound(P, P') \cap \lpref{P'}$.
The following theorem gives a fixpoint characterization of $\rwamay$.
\begin{theorem}[\cite{lha_techrep11}] \label{thm:may} 
  For all locations $l$ and polyhedra $U$, $V$, it holds
%  $\tau(U,V,W) =$
%
{%\small
  \begin{equation} \label{eq:our-fixp}
   \rwamay_l(U, V) = \mu W \qdot  
    U \cup
    \bigcup_{P\in\conv{\overline{V}}} \bigcup_{P'\in \conv{W}} \Bigl(
    P \cap \lpref{entry(P,P')} \Bigr).
  \end{equation}
}
\end{theorem}

\begin{figure}[h!]
\centering
%\subfigure[Initial input.]{
%\label{fig:removeunsafe}
%\includegraphics[width=1.3in]{fig/RemoveUnsafe.pdf}
%}
\subfigure[Initial input, with $\bound(P, P')$ highlighted.]{
\label{fig:removeunsafe1}
\includegraphics[width=1.2in]{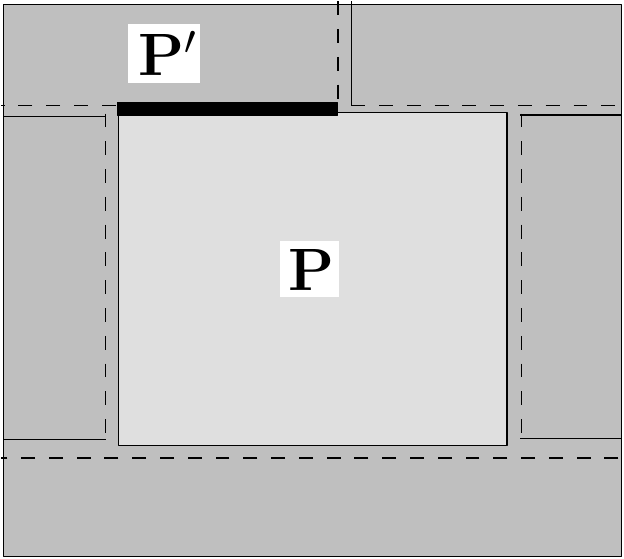}
}
\subfigure[Pre-flow of $P'$.]{
\label{fig:removeunsafe2}
\includegraphics[width=1.4in]{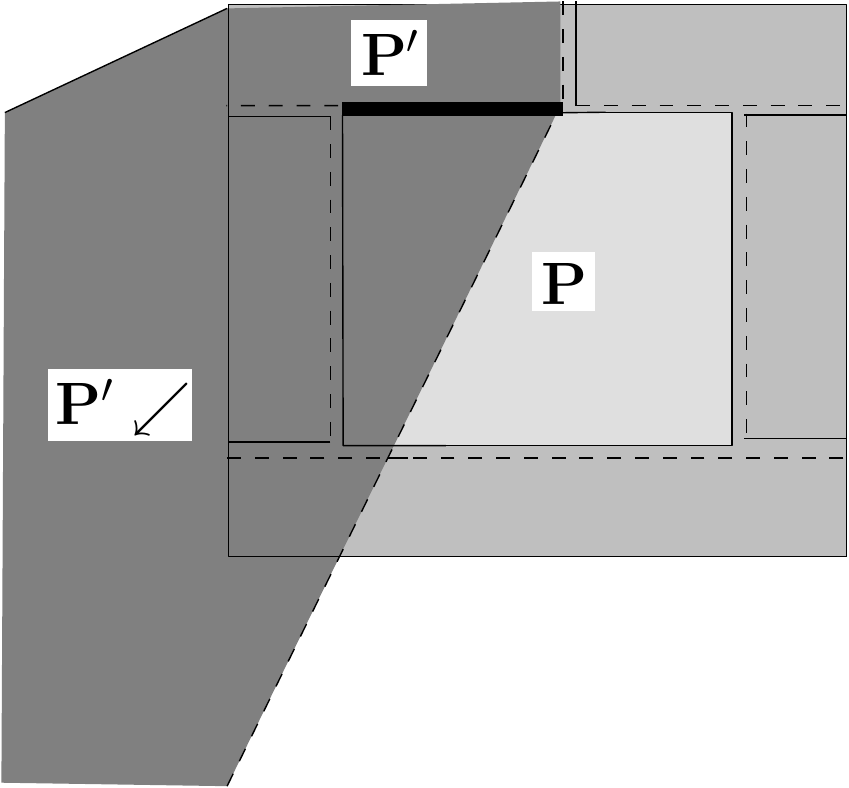}
}
\subfigure[Entry region.]{
\label{fig:removeunsafe3}
\includegraphics[width=1.4in]{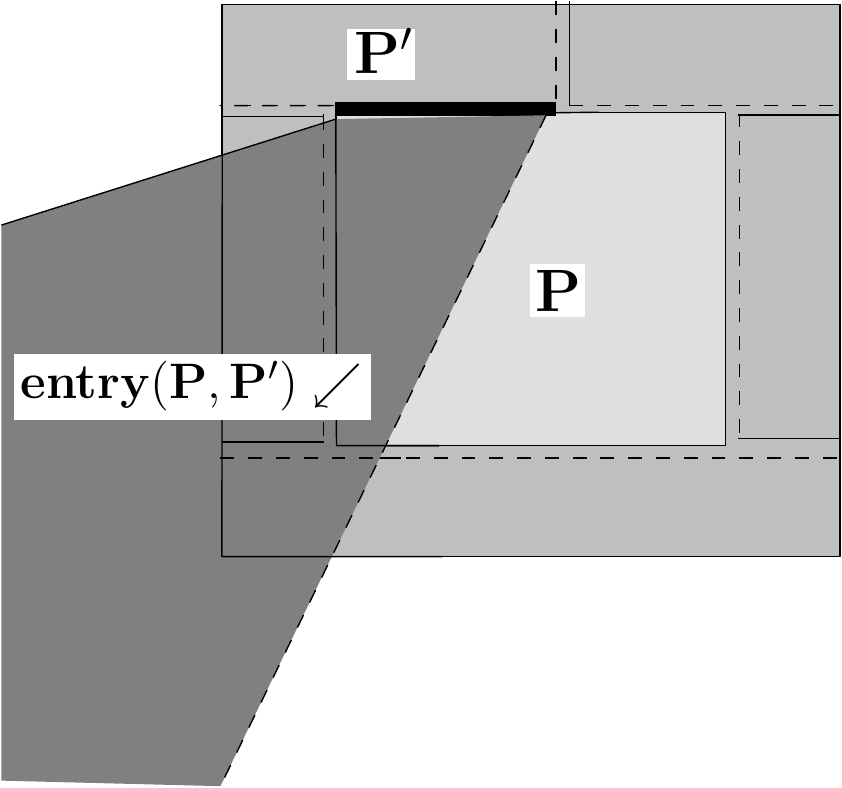}
}
\subfigure[$\pnew$, $\cut$.]{
\label{fig:removeunsafe4}
\includegraphics[width=1.2in]{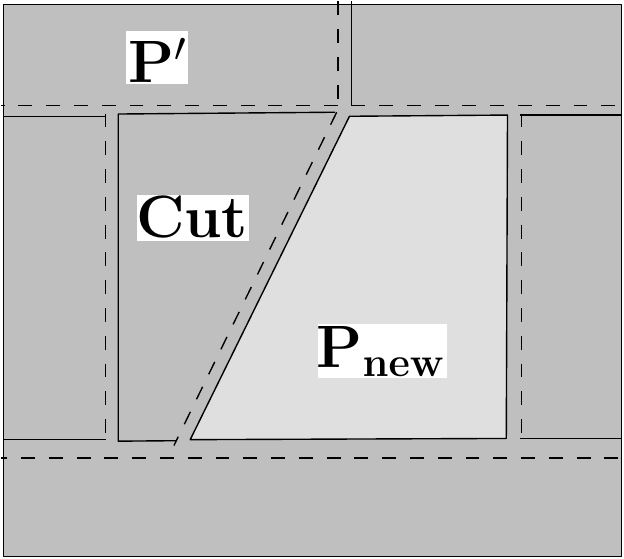}
}
\subfigure[$\Flow(l)$.]{
\label{fig:removeunsafe5}
\includegraphics[width=0.5in]{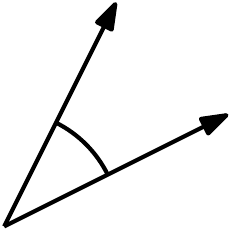}
}
\caption{Algorithm behavior.}
\label{fig:tanks}
\end{figure}

The equation refines the under-approximation $U$ by identifying its
\emph{entry regions}, i.e., the boundaries between the area which
\emph{may} belong to the result (i.e., $\overline{V}$), and the area
which already belongs to it (i.e., $W$).
%
%For all $P\in\conv{\overline{V}}$ and $P' \in\conv{W}$, let
%$b=\bound(P,P')$, and $entry(P,P') = b \cap \lpref{P'}$.  
%The set
%$entry(P,P')$, that we call the entry region \emph{from $P$ to $P'$},
%contains the points of $b$ that may reach $P'$ by following a
%trajectory of the system.
%and also an entry region \emph{of $W$}.
%Hence, the system may move from $P$ to $P'$ through $entry(P,P')$.
Figure~\ref{fig:tanks} shows a single step in the computation of
equation~\ref{eq:our-fixp}, for a fixed pair of convex polyhedra $P$
in $\overline{V}$ and $P'$ in $W$. 
Dashed lines represent topologically open sides.
The dark gray rectangles represent
convex polyhedra in $W$, while the light gray one is $P$.

In Figure~\ref{fig:removeunsafe1} the thick segment between $P$ and
$P'$ represents $\bound(P,P')$ and, in the example, is contained in $P$.
Since $P'$ is topologically open (denoted by the dashed line),
the rightmost point of $\bound(P,P')$ cannot reach $P'$ along any
straight-line activity.
%shows the set $entry(P,P')$ where
%the points of $\bound(P,P')$ which cannot reach $P'$ have been
%discarded. 
Being $P'$ open, so is $\lpref{P'}$, and its intersection with $P$,
namely $entry(P,P')$, does not contain the rightmost point of the
boundary (Figure~\ref{fig:removeunsafe2}). Now, any point of $P$ that
can reach $entry(P,P')$ following some activity can also reach $P'$,
and
%The set of points
%$\lpref{entry(P,P')}$, which can reach $entry(P,P')$, and hence $P'$,
%following some activity, is shown in Figure~\ref{fig:removeunsafe3}.
the set $\cut = P \cap \lpref{entry(P,P')}$ contains precisely those
points (Figure~\ref{fig:removeunsafe3} and
Figure~\ref{fig:removeunsafe4}). All these points must then be added
to $W$, as they all belong to $\rwamay_l(U, V)$.

In our implementation, instead of computing the operator $\rwamay_l$, we
compute the dual operator $\staymust_l(Z,V)$ (for \emph{must stay or reach}),
containing the points which
either remain in $Z$ forever or reach $V$ along a system trajectory
that does not leave $Z$.
The operator $\staymust_l$ can be defined as follows:
\begin{equation} \label{eq:sorm-def}
\staymust_l(Z,V) = \overline{\rwamay_l(\overline{Z},V)}.
\end{equation}
As a consequence, we can compute $\cpre(A)$ as
{\small
$$
\bigcup_{l \in \Loc} \{ l \} \times \Big( \proj{A}{l} \cap
\staymust_l\big( \proj{\overline{\Inv}}{l} \cup \big( \proj{A}{l}\!\setminus B_l \big),
               C_l \cup \overline{\proj{\Inv}{l}} \big) \Big).
$$
}
From~\eqref{eq:sorm-def}, we obtain a fixpoint characterization of the
operator $\staymust_l$:
{\small

\begin{align}
\staymust_l(Z,V)&=\overline{\rwamay_l(\overline{Z},V)} 
= \overline{\mu W \qdot \overline{Z} \cup
  \bigcup_{P\in\conv{\overline{V}}} \bigcup_{P'\in \conv{W}}
  \big( P \cap \lpref{entry(P,P')} \big)} =\notag\\[1ex]
&= \nu W \qdot Z \cap
\overline{\bigcup_{P\in\conv{\overline{V}}} \bigcup_{P'\in \conv{\overline{W}}}
\big( P \cap \lpref{entry(P,P')} \big)}  
= \nu W \qdot Z \setminus
\bigcup_{P\in\conv{\overline{V}}} \bigcup_{P'\in \conv{\overline{W}}}
\big( P \cap \lpref{entry(P,P')} \big). \label{eq:sorm-fixp}
\end{align}
}
The following two sections show how to effectively and efficiently
compute fixpoint~\eqref{eq:sorm-fixp}.
%
%We can interpret the last line as an incremental refinement 
%of the over-approximation $Z$ of $\staymust_l(Z, V)$ to the desired
%result.
%

%Using the definition of $\cpre_l(A)$ in term of $\staymust_l$ operator, we have to compute $\staymust_l(Z,V)$, where $Z=\proj{\overline{\Inv}}{l} \cup \big( \proj{A}{l} \setminus B_l \big)$ and $V=C_l \cup \overline{\proj{\Inv}{l}}$.

\section{Exact Computation of Pre-Flow} \label{sec:preflow}

As seen in the previous section, one of the basic operations
on polyhedra that are needed to compute $\staymust$ is the pre-flow
operator $\preftag$. It is sufficient to compute $\pref{P}{F}$ for
convex $P$ and $F$, for two reasons:
First, we always have $F = \Flow(l)$,
for a given location $l$, and $\Flow(l)$ is a convex polyhedron by assumption.
Second, $\pref{(P_1 \cup P_2)}{F} = (\pref{P_1}{F}) \cup (\pref{P_2}{F})$,
so the pre-flow of a general polyhedron is the union of the pre-flows
of its convex polyhedra. \\
The pre-flow of $P$ w.r.t. $F$ is equivalent to the \emph{post-flow}
of $P$ w.r.t. $-F$, defined as:
$$
\postf{P}{-F} = \{ x + \delta \cdot y \mid x\in P, y\in -F, \delta\geq 0 \}.
$$
The post-flow operation coincides with the \emph{time-elapse}
operation introduced in~\cite{halbwachs97} 
for topologically closed convex polyhedra.
Notice that for convex polyhedra $P$ and $F$, 
the post-flow of $P$ w.r.t. $F$ may not be a convex polyhedron:
following~\cite{asarin00}, let $P \subseteq \reals^2$
be the polyhedron containing only the origin $(0,0)$ and let
$F$ be defined by the constraint $y>0$.
We have $\postf{P}{F} = \{ (0,0) \} \cup \{ (x,y) \in \reals^2 \mid y>0 \}$,
which is not a convex polyhedron 
(although it is a convex subset of $\reals^2$).
The Parma Polyhedral Library (PPL, see~\cite{ppl08}), for instance,
only provides an over-approximation $\postfppl{P}{F}$
of the post-flow $\postf{P}{F}$, as
the smallest convex polyhedron containing $\postf{P}{F}$.

On the other hand, the post-flow of a convex polyhedron is always the union of
two convex polyhedra, according to the equation
$$
\postf{P}{F} = P \cup \big(\pospostf{P}{F}\big),
$$
where $\pospostf{P}{F}$ is the \emph{positive post-flow} of $P$,
i.e., the set of valuations that can be reached from $P$
via a straight line of non-zero length whose slope belongs to $F$. Formally,
$$
\pospostf{P}{F} = \{ x + \delta \cdot y \mid x\in P, y\in F, \delta>0 \}.
$$
Hence, in order to exactly compute the post-flow of a convex polyhedron,
we show how to compute the positive post-flow.

Convex polyhedra admit two finite representations, in terms of
\emph{constraints} or \emph{generators}.
Libraries like PPL maintain both representations for each convex polyhedron
and efficient algorithms exist for keeping them synchronized
\cite{chernikova68,verge92}.
The constraint representation refers to the set of linear inequalities 
whose solutions are the points of the polyhedron.
The generator representation consists in three finite sets of 
\emph{points}, \emph{closure points}, and \emph{rays},
that generate all points in the polyhedron by linear combination.
More precisely, for each convex polyhedron $P \subseteq \reals^n$
there exists a triple $(V,C,R)$ such that $V$, $C$, and $R$ are finite sets
of points in $\reals^n$, and $x \in P$ if and only if it can be written
as
\begin{equation} \label{eq:combination} 
\sum_{v \in V} \alpha_v \cdot v + \sum_{c \in C} \beta_c \cdot c +
\sum_{r \in R} \gamma_r \cdot r,
\end{equation}
where all coefficients $\alpha_v$, $\beta_c$ and $\gamma_r$ are non-negative
reals, $\sum_{v \in V} \alpha_v + \sum_{c \in C} \beta_c = 1$,
and there exists $v \in V$ such that $\alpha_v > 0$.
We call the triple $(V,C,R)$ a \emph{generator} for $P$. \\
Intuitively, the elements of $V$ are the proper vertices of the polyhedron $P$,
the elements of $C$ are vertices of the topological closure of $P$ that
do not belong to $P$,
and each element of $R$ represents a direction of unboundedness of $P$.

The following result shows how to efficiently compute the positive post-flow
operator, using the generator representation.
\begin{theorem}
Given two convex polyhedra $P$ and $F$,
let $(V_P,C_P,R_P)$ be a generator for $P$ and $(V_F,C_F,R_F)$
a generator for $F$.
The triple $(V_P\oplus V_F, C_P \cup V_P, R_P \cup V_F \cup C_F \cup R_F)$
is a generator for $\pospostf{P}{F}$, where $\oplus$ denotes Minkowski sum.
\end{theorem}
%\begin{proof}
\textbf{Proof} %Provv
Let $z \in \pospostf{P}{F}$, we show that there are coefficients 
$\alpha_v$, $\beta_c$ and $\gamma_r$ such that $z$ can be written 
as~\eqref{eq:combination}, 
for $V=V_P \oplus V_F$, $C=C_P \cup V_P$, and $R=R_P \cup V_F \cup C_F \cup R_F$.

By definition, there exist $x \in P$, $y \in F$, and $\delta>0$ such that
$z = x + \delta y$.
Hence, there are coefficients 
$\alpha^x_v$, $\beta^x_c$, and $\gamma^x_r$ witnessing the fact that $x \in P$,
and coefficients
$\alpha^y_v$, $\beta^y_c$, and $\gamma^y_r$ witnessing the fact that $y \in F$.
Moreover, there is $i \in V_P$ and $j \in V_F$ such that $\alpha^x_i >0$
and $\alpha^y_j >0$.
Let $\varepsilon = \min \{\alpha^x_i, \delta \alpha^y_j \}$
and notice that $\varepsilon > 0$.
It holds 
\begin{align*}
\alpha^x_i \cdot i + \delta \cdot \alpha^y_j \cdot j &=
(\alpha^x_i - \varepsilon) i + \varepsilon i +
(\delta \cdot \alpha^y_j  - \varepsilon) j +
\varepsilon j 
= \varepsilon (i+j) + 
(\alpha^x_i               - \varepsilon) i +
(\delta \cdot \alpha^y_j  - \varepsilon) j.
\end{align*}
Hence, 
\begin{align*}
z &= \sum_{v \in V_P} \alpha^x_v \cdot v + 
     \sum_{c \in C_P} \beta^x_c  \cdot c +
     \sum_{r \in R_P} \gamma^x_r \cdot r 
   + %\\
     \quad\delta \Bigg(\sum_{v \in V_F} \alpha^y_v \cdot v + 
     \sum_{c \in C_F} \beta^y_c  \cdot c +
     \sum_{r \in R_F} \gamma^y_r \cdot r \Bigg)
\\
  &= \varepsilon (i+j) + 
  \quad \Bigg( (\alpha^x_i - \varepsilon) i +
       \sum_{v \in V_P\setminus \{ i\}} \alpha^x_v \cdot v + 
       \sum_{c \in C_P} \beta^x_c \cdot c \Bigg) + \\
  &\quad \Bigg( (\delta \cdot \alpha^y_j  - \varepsilon) j +
       \sum_{r \in R_P} \gamma^x_r \cdot r +
       \sum_{v \in V_F\setminus \{ j \}} \alpha^y_v \cdot v + 
  \quad \sum_{c \in C_F} \beta^y_c  \cdot c +
         \sum_{r \in R_F} \gamma^y_r \cdot r \Bigg).
\end{align*}
One can easily verify that: \emph{(i)} all coefficients are non-negative;
\emph{(ii)} the sum of the coefficients of the points in $V$ and $C$ is $1$;
\emph{(iii)} there exists a point in $V$, namely $i+j$, such that its
coefficient is strictly positive.

Conversely, let $z$ be a point that can be expressed as~\eqref{eq:combination}, 
for $V=V_P \oplus V_F$, $C=C_P \cup V_P$, and $R=R_P \cup V_F \cup C_F \cup R_F$.
We prove that $z \in \pospostf{P}{F}$ by identifying $x\in P$,
$y\in F$ and $\delta>0$ such that $z = x + \delta y$. \\
Notice that \emph{(a)} $\sum_{v \in V_P \oplus V_F} \alpha_v + \sum_{c
  \in C_P \cup V_P} \beta_c =1$, and \emph{(b)} there exists $v^* \in
V_P \oplus V_F$ such that $\alpha_{v^*} > 0$.  We set
$$
x = \sum_{\substack{v_1 \in V_P\\v_2 \in V_F}} \alpha_{v_1+v_2} \cdot v_1 +
    \sum_{c \in C_P \cup V_P} \beta_c \cdot c +
    \sum_{r \in R_P} \gamma_r \cdot r.
$$
We claim that $x \in P$: first, $x$ is expressed as a linear
combination of points in $(V_P, C_P, R_P)$;
second, all coefficients are non-negative;
third, the sum of the coefficients of the points in $V_P$ and in $C_P$ is $1$,
due to \emph{(a)} above; finally, since $\alpha_{v^*} > 0$,
there is a point in $V_P$ whose coefficient is positive.
Then, we set
\begin{align*}
\delta = \sum_{v \in V_P \oplus V_F} \alpha_v + \sum_{r \in V_F \cup C_F} \gamma_r,
\qquad \text{ and }\qquad 
y = \frac{1}{\delta} \cdot \Bigg(
   \sum_{\substack{v_1 \in V_P\\v_2 \in V_F}} \alpha_{v_1+v_2} \cdot v_2 +
   \sum_{r \in V_F \cup C_F \cup R_F} \gamma_r \cdot r \Bigg).
\end{align*}
Since $\alpha_{v^*} > 0$, we have $\delta > 0$.
We claim that $y \in F$: first, $y$ is a linear
combination of points in $(V_F, C_F, R_F)$;
second, all coefficients are non-negative;
third, the sum of the coefficients of the points in $V_F$ and in $C_F$ is $1$,
due to our choice of $\delta$; finally, since $\alpha_{v^*} > 0$,
there is a point in $V_F$ whose coefficient is positive.
%\end{proof}

\section{Computing $\staymust$} \label{sec:sorm}

\def\bps{B}
\def\res{\mathit{Res}}
\def\checkint{\ensuremath{\mathit{UpdInt}}}
\def\checkext{\ensuremath{\mathit{UpdExt}}}
\def\refine{\ensuremath{\mathit{Refine}}}
\def\initentry{\ensuremath{\mathit{PotentialEntry}}}
\def\cand{\ensuremath{\mathit{Candidates}}}

In this section, we show how to efficiently compute $\staymust_l(Z,V)$,
given two polyhedra $Z$ and $V$.
Fixpoint equation~\eqref{eq:sorm-fixp} can easily be converted into an iterative
algorithm, consisting in generating a (potentially infinite) 
sequence of polyhedra $(W_n)_{n \in \nats}$,
where $W_0=Z$ and
\begin{equation} \label{eq:step1}
W_{i+1} = W_i \setminus \bigcup_{P\in\conv{\overline{V}}} 
                       \bigcup_{P'\in \conv{\overline{W_i}}}
                           \Bigl( P \cap \lpref{entry(P,P')}\Bigr).
\end{equation}
Theorem~4 in~\cite{lha_techrep11} proves that such sequence converges
to a fixpoint within a finite number of steps.
%The fixpoint refines the overapproximation $Z$, 
%When the fixpoint is reached ($W_n=W_{n-1}$), no more area can be removed from 
%$W_{n-1}$ and $W_{n-1}$ is the result of $\staymust_l(Z,V)$.
%
The naive implementation of the algorithm is done by an outer loop
over the polyhedra $P\in\conv{\overline{V}}$ and an inner loop over
$P'\in\conv{\overline{W_i}}$.  As a first improvement, we notice that
each iteration of the outer loop removes from $W_i$ a portion of
$P\in\conv{\overline{V}}$.  Hence, the portion of $P$ that is not
contained in $W_i$ is irrelevant, and we may replace~\eqref{eq:step1}
with:
\begin{equation} \label{eq:step2} W_{i+1} = W_i \setminus
  \bigcup_{P\in\conv{W_i \cap\overline{V}}} \bigcup_{P'\in
    \conv{\overline{W_i}}} \Bigl( P \cap \lpref{entry(P,P')}\Bigr).
\end{equation}
Moreover, we can avoid the need to intersect $W_i$ with $\overline{V}$
at each iteration, by starting with $W_0' = Z \setminus V$,
setting:
\begin{equation} \label{eq:step3}
W'_{i+1} = W'_i \setminus \bigcup_{P\in\conv{W'_i}} 
                       \bigcup_{P'\in \conv{\overline{W'_i}}}
                           \Bigl( P \cap \lpref{entry(P,P')}\Bigr),
\end{equation}
and noticing that $W_i = W'_i \cup V$ for all $i\geq 0$.
As a consequence, $\staymust_l(Z,V) = \lim_{i\to\infty} W_i = 
V \cup \lim_{i\to\infty} W'_i$.
The implementation described so far is called the \emph{basic
  approach} in the following.

\subsection{Introducing Adjacency Relations} \label{sec:adj}

\begin{comment}
\begin{algorithm}[ht]
 %\dontprintsemicolon
\KwIn{Set of $\nnc$ pairs $\int$, $\ext$,
$\PS$ $V$, $\nnc$ $F$}
\KwOut{$\PS$ $\res$}
%\KwData{hashmap $\int$, $\ext$, list $\queue$, Polyhedra $V$, convex polyhedron $F$.}
  \While{$\queue\neq\emptyset$}{
    $P    \leftarrow \queue.dequeue()$\;
    $\bps \leftarrow \bigcup \big\{R \mid \pair{P,R}\in\ext\big\}$\;
    $\cut \leftarrow P \cap (\lpref{\bps})$\;
    \If{$\cut\neq\emptyset$}{
       $\pnew\leftarrow P\setminus\cut$\;
       \ForEach{$P'\in \conv{\pnew}$}{
         $\checkint(\int, P', \pnew)$\;
       }
       \ForEach{$P'\in \{Q \mid \pair{P, Q}\in\int\wedge Q\not\subseteq V\}$}{
         
         $\checkint(\int, P', \pnew)$\;
         $\checkext(\ext, \queue, P', \cut, F, V)$\;
	 % was: $P_{entry}\leftarrow entry\_regions(P',cut,F)$\;
	 %      $\checkext(\ext, P', P_{entry}, F, V, \queue)$\;
       }
       $\int\leftarrow\int\setminus\{\langle P, X\rangle\in\int\}$\;
       $\ext\leftarrow\ext\setminus\{\langle P, X\rangle\in\ext\}$\;       
     }
  }
  \Return $\{P \mid \langle P, P'\rangle\in\int\}$\;
\caption{$\refine(\int, \ext, \queue, V)$}\label{algo:RefineMap}
\end{algorithm}
\end{comment}

Given two disjoint convex polyhedra $P$ and $P'$, we say that they are
\emph{adjacent} if $\bound(P,P') \neq \emptyset$.  In the basic
approach, the inner loop is repeated for each
$P'\in\conv{\overline{W_i}}$, even if convex polyhedra $P'$ that are
not adjacent to $P$ result in an empty $entry(P,P')$ and are therefore
irrelevant.  Hence, we define the binary relation of \textit{external
  adjacency} $\ext_i$, which associates a polyhedron $P\in\conv{W_i}$
with its entry regions $entry(P,P')\neq\emptyset$, for all
$P'\in\conv{\overline{W}_i}$.  Formally,
\begin{align} \label{eq:ext}
\ext_i = \big\{ \pair{P,entry(P,P')} \mid P\in\conv{W_i},
P'\in\conv{\overline{W}_i},
\text{ and  } entry(P,P')\neq\emptyset \big\}.
\end{align}
Once $\ext_i$ is introduced and properly maintained, it also enables
to optimize the outer loop.  Rather than $P\in\conv{W_i}$, it is
enough to consider all $P$ which are associated with at least one
entry region in $\ext_i$, i.e., all $P$ such that
$\pair{P,R}\in\ext_i$ for some $R$.
Summarizing, using $\ext_i$ we can replace~\eqref{eq:step3} with
\begin{equation} \label{eq:step4}
W_{i+1} = W_i \setminus \bigcup_{\pair{P,R} \in \ext_i} 
                       \Bigl( P \cap \lpref{R}\Bigr).
\end{equation}

\begin{comment}
Notice that: 
\begin{enumerate}
\item $\vert\{P\vert\langle P,R\rangle\in\ext_0$, for some
  $R\}\vert\leq\vert\conv{W_1\cap\overline{V}}\vert$, and
\item $\vert\{R\vert\langle
  P,R\rangle\in\ext_0\}\vert\leq\vert\conv{\overline{W}_1}\vert$.
\end{enumerate}
%
This means that using the external adjacency, we can reduce the number
of cycles needed in the second step of the main loop.
\end{comment}

Clearly, some extra effort is required to initialize and maintain
$\ext_i$.  Initialization is performed by simply
applying~\eqref{eq:ext}.  Regarding maintenance, we briefly discuss
how to efficiently compute
$\ext_{i+1}$. \\

\noindent\begin{minipage}[b]{.53\textwidth}
\small\begin{algorithm}[H]
  \KwIn{$\PS$ $Z$, $V$, $\nnc$ $F$}
  \KwOut{$\PS\ \staymust(Z, V, F)$}
  \ForEach{$\nnc$ $P\in\conv{Z}$}{
    $\intnew \leftarrow \checkint(\intnew, P, Z)$\;
    $E \leftarrow \initentry(P, \intnew,F)$\;
    $\extnew \leftarrow \checkext(\extnew, P, E, F, V)$\;
  }
  \While{$\extnew\neq\emptyset$} {
    $\extold \leftarrow \extnew$\;
    $\intold \leftarrow \intnew$\;
    $\extnew\leftarrow\emptyset$\; 
    \ForEach{$P$ s.t. $\pair{P,R}\in\extold$} {
      $\bps \leftarrow \bigcup \big\{R \mid \pair{P,R}\in\ext_i\big\}$\;
      $\cut \leftarrow P \cap (\lpref{\bps})$\;
      \If{$\cut\neq\emptyset$}{
        $\pnew\leftarrow P\setminus\cut$\;
        \ForEach{$P'\in \conv{\pnew}$} {
          $\intnew \leftarrow \checkint(\intnew,P', \pnew)$\;
        }
        \ForEach{$P'$ s.t. $\pair{P, P'}\in\intold$}{             
          $\intnew \leftarrow \checkint(\intnew, P', \pnew)$\;
          \mbox{$\extnew \leftarrow \checkext(\extnew, P', \cut, F, V)$\;}
        }
        $\intnew\leftarrow\intnew\setminus\{\langle P, Q\rangle\in\intold\}$\;
      }
    }
  }
  \Return $\{P \mid \langle P, P'\rangle\in\intnew\}$\;
  \caption{$\staymust(Z, V, F)$}\label{algo:sorm}
\end{algorithm}
\end{minipage}
\begin{minipage}{.47\textwidth}\small
  \begin{algorithm}[H]
\vspace{2pt}
   \KwIn{Set of $\nnc$ pairs $\int$; $\nnc$ $P$;\\ \hspace{30pt} $\PS$ $\cand$;}
   \KwOut{Set of $\nnc$ pairs $\int$;}
\vspace{4pt}
    $\int\leftarrow \int \cup \{ \pair{P,\emptyset} \}$\;	  
    \ForEach{$\nnc$ $P'\in \conv{\cand}$, with $P'\neq P$}{
      \If{$\bound(P, P')\neq\emptyset$}{
        $\int\leftarrow\int\cup\{\langle P, P'\rangle\}$\;	  
      }
    }
    \Return $\int$\;
    \caption{$\checkint(\int, P, \cand)$}\label{algo:updint}
    \vspace{2pt}  
\end{algorithm}
\vspace{9pt}
  \begin{algorithm}[H]
\vspace{2pt}
    \KwIn{Set of $\nnc$ pairs $\ext$; $\nnc$ $P, F$; \\
     \hspace{30pt}$\PS$ $\cand, V$;}
    \KwOut{Set of $\nnc$ pairs $\ext$;}
\vspace{4pt}
    \If{$P\not\subseteq V$}{
      \ForEach{$\nnc$ $P'\in \conv{\cand}$}{
        $R\leftarrow entry(P, P')$\; 
        \If{$R\neq\emptyset$}{
          $\ext\leftarrow\ext\cup\{\langle P, R\rangle\}$\;
        }
      }
    }
    \Return $\ext$\;  
\vspace{2pt}
 \caption{$\checkext(\ext, P, \cand, F, V)$}\label{algo:updext}
\end{algorithm}
\end{minipage}
\vspace{6pt}

During the $i$-th iteration, certain convex polyhedra $P \in
\conv{W_{i}}$ are cut by removing the points that may directly reach a
convex polyhedron $P' \in \conv{\overline{W}_i}$.  These cuts may
\emph{expose} other convex polyhedra in $\conv{W_{i}}$, that were
previously covered by $P$.  These exposed polyhedra will be the only
ones to have associated entry regions in $\ext_{i+1}$.  In order to be
exposed by a cut made to $P$, a convex polyhedron must be adjacent to
$P$.  Hence, in order to compute $\ext_{i+1}$ it is useful to have
information about the adjacency among the polyhedra in $\conv{W_{i}}$.
To this aim, we also introduce the binary relation of \emph{internal
  adjacency} $\int_i$ between polyhedra in $\conv{W_i}$:
\begin{align}
\int_i = \big\{ \pair{P_1, P_2} \mid P_1, P_2\in\conv{W_i},
P_1\neq P_2
\text{ and } \bound(P_1,P_2)\neq\emptyset \big\}.
\end{align}
The computation of $\int_0$ requires the complete scan of all
$P_1,P_2\in\conv{W_0}$, while $\int_{i+1}$ is obtained incrementally
from $\int_i$ and $\ext_i$.  Given $\pair{P, R} \in \ext_i$, let $\cut
= P \cap \big(\lpref{R}\big)$ and $\pnew = P \setminus \cut$.  Notice
that $\pnew$ may be non-convex, being the result of a set-theoretical difference
between two convex polyhedra.
To obtain $\int_{i+1}$, we add to $\int_i$ the pairs of adjacent
convex polyhedra $(P_1,P_2)$ such that either \emph{(i)} both $P_1$
and $P_2$ belong to $\conv{\pnew}$, or \emph{(ii)} one of them belongs
to $\conv{\pnew}$ and the other is adjacent to $P$ according to
$\int_i$.  Moreover, once $\pnew$ replaces $P$ in $W_{i+1}$, it is
necessary to remove all the pairs $\pair{P,P'}$ from $\ext_i$ and
$\int_i$.

%
% Notice that, in the $(i+1)$-th step, the only $Q\in\conv{W_{i+1}}$
% that may be associated with a new entry region, are those that in
% the $i$-th step were adjacent to a polyhedron $P\in\conv{W_{i}}$
% which was removed the area $\cut = P\cap\lpref(R)$, where
% $\pair{P,R}\in\ext_{i}$.  Cleary, the only new entry regions are
% those generated by $\cut$.  Hence, then computing the entry region
% from $Q$ to $\cut$ that is
%
% $$\ext_{i+1} = \ext_{i} \cup \{ \pair{Q, entry(Q,\cut)} \}$$
%
% for each $P'$ adjacent to $P$ and such that
% $entry(P',cut)\neq\emptyset$.

Algorithms~\ref{algo:sorm}-\ref{algo:updext} represent a concrete
implementation of the technique described so far.  In
Algorithm~\ref{algo:sorm}, $\extold$ and $\intold$ represent the old
adjacency relations, while $\extnew$ and $\intnew$ the new ones.  The
first ``for each'' loop initializes both relations, followed by a
``while'' loop that iterates until the external adjacency relation is
empty.  Maintenance of the adjacency relations is delegated to
Algorithms \ref{algo:updint} and \ref{algo:updext}, that receive as
input the relation they have to update, the convex polyhedron $P$
whose adjacencies need to be examined, and a general polyhedron
$\cand$ containing the convex polyhedra that may be adjacent to $P$.
Additionally, Algorithm \ref{algo:updext} also needs to know the input
set $V$ (region to be avoided) and the location flow $F = \Flow(l)$.

The auxiliary function $\initentry$ returns the potential entry region for $P$.
In this version, we simply have
$$
\initentry(P, \int_0, F) = \bar{Z}.
$$
This will be improved in Section~\ref{sec:further}.

\subsection{Further Improving the Performance} \label{sec:further}

Recall that $\initentry(P,\int_0,F)$ returns $\bar{Z}$, regardless of
its inputs. Experimental evidence (see Section~\ref{sec:micro})
shows that it is often the case that
the portion of $\bar{Z}$ which is relevant to computing the entry
regions of a given a convex polyhedron $P$ is much smaller than the whole
set $\bar{Z}$. This often leads to a large number of attempts to
compute entry regions which end up empty.
%
%
% computing the complement of $Z$ sometimes becomes a bottleneck for
% the whole algorithm, as explained in Section~\ref{sec:experiments}.
% To avoid this, we present a technique that
To avoid this, for each $P$ in $\conv{Z}$ we proceed as follows.  We first
collect $P$ and all convex polyhedra in $\conv{Z}$ that are adjacent to it:
$P_{adj}=\{P\} \cup \{P' \mid \pair{P, P'} \in\int_0 \}$.  Then, we
compute
$$
\initentry(P, \int_0, F) = (P\nearrow F)\setminus P_{adj}.$$ 
The resulting polyhedron contains all and only the convex polyhedra of
$\bar{Z}$ which, if adjacent to $P$, give rise to a non-empty entry
region.
% This manages the fact that the entry regions can be only in $V$,
% without considering all $\overline{Z}$.  In a similar manner, the
% function $entry\_region(P, cut, F)$ called by Algorithm
% \ref{algo:RefineMap}, return the area $external=(P\nearrow F)\cap
% cut$.

\section{Experiments with PHAVer+} \label{sec:experiments}

We implemented the three algorithms described in the previous section on the top of the open-source tool PHAVer~\cite{phaver05}.
In the following figures,
the basic approach (Section~\ref{sec:sorm}) is denoted by \emph{Basic},
the adjacency approach (Section~\ref{sec:adj}) by \emph{Adj}, and
the local adjacency approach (Section~\ref{sec:further}) by \emph{Local}.
We show some results obtained by testing our package 
on two different examples: the Truck Navigation Control (TNC) 
and the Water Tanks Control (WTC).
The experiments are divided into two distinct categories:
the \emph{macro} analysis shows the performance of the three implementations
when solving safety control problems,
% (also compared with the performance achieved by {\sc HoneyTech}).
while the \emph{micro} analysis shows the performances of a single call to
the $\staymust_l(Z,V)$ operator.
%, implemented in the three different ways presented earlier.
%, showing the trend compared to the size of the input (which will be defined below): this will help us better understand the reasons for greater speed of execution of an implementation with respect to other. 
A binary pre-release of our implementation, that we call PHAVer+, can be downloaded at {\tt http://people.na.infn.it/mfaella/phaverplus}.
The experiments were performed on an Intel Xeon (2.80GHz) PC.

\subsection{Macro Analysis}
We now describe in detail the two examples used to evaluate
the performance of our package.

\begin{figure}
\begin{center}
\includegraphics[width=3.5in]{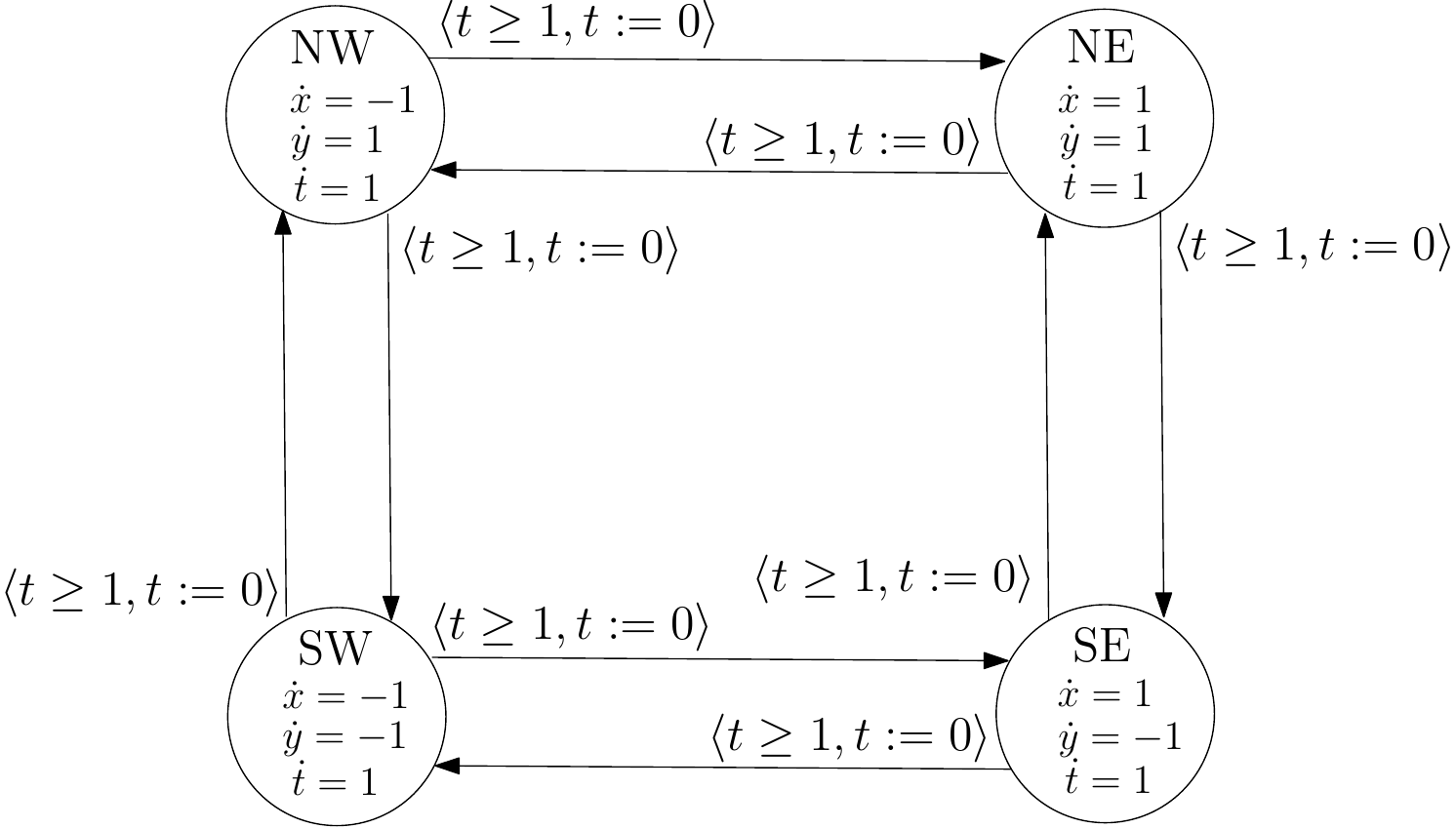}
\end{center}
\caption[TNC modeled as a Hybrid Automaton.]{TNC modeled as a Hybrid Automaton.}\label{fig:TNC}
\end{figure}

\paragraph{Truck Navigation Control.}
This example is derived from~\cite{honeytech}, where the tool {\sc HoneyTech} is presented,
as an extension of {\sc HyTech}~\cite{hytech97} for the automatic synthesis of controllers.
Consider an autonomous toy truck, which is responsible for avoiding some 2 by 1 rectangular pits. 
The truck can take 90-degree left or right turns: the possible directions are North-East (NE), 
North-West (NW), South-East (SE) and South-West (SW). One time unit must pass between two changes of  direction. The control goal consists in avoiding the pits.
%The corresponding hybrid automaton has four locations, one for each direction, and three variables: $x$ and $y$ %for the truck position, and $t$ as the clock that enforces a one-time-unit wait between consecutive changes of %direction.
%
%Notice that the TNC proposed in~\cite{honeytech} is limited to one turn only, while our analysis is extended to the complete case (an unlimited number of turns is allowed).
Figure~\ref{fig:TNC} shows the hybrid automaton modeling the system: there is one location for each direction, where the derivative of the position variables ($x$ and $y$) are set according to the corresponding direction. The variable $t$ represents a clock ($\dot{t}=1$) that enforces a one-time-unit wait between turns.

\begin{figure}[h!]
\centering
\subfigure[Performance for TNC.]{
\includegraphics[width=2.0in]{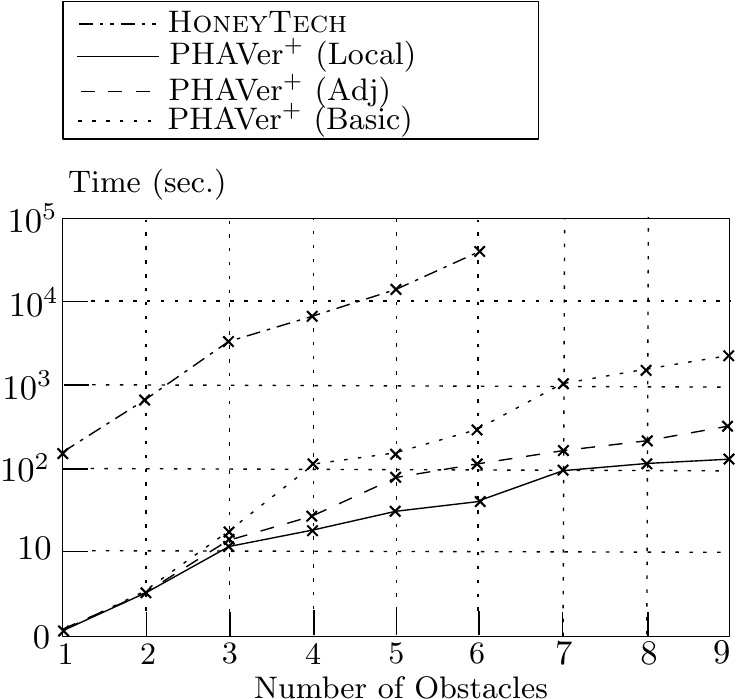}
\label{fig:perf}
}
\subfigure[System schema for WTC.]{
\includegraphics[width=2.0in]{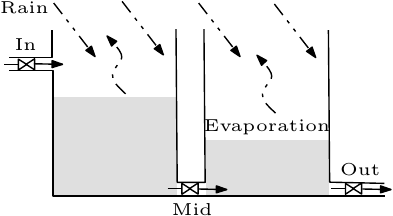}
\label{fig:TanksDescr}
}
\subfigure[Performance for WTC.]{%
\begin{tabular}[b]{c|c}
    \hline
    Algorithm  &  Time (sec.) \\ \hline\hline
    Basic      & 21.0 \\ \hline
    Adj        & 16.2 \\ \hline
    Local      & 9.3  \\ \hline
\end{tabular}
\label{fig:TanksTime}
}
\caption{Schema and performance for the two examples.}
\label{fig:tanksglobal}
\end{figure}

We tested our implementations on progressively more complex control goals,
by increasing the number of obstacles. Figure~\ref{fig:perf} compares
the performance of the three implementations of the algorithm (solid line for local,
dashed line for adjacency, dotted line for basic and dotted-dashed line for the performance reported in~\cite{honeytech}). We were not able to replicate the experiments in~\cite{honeytech}, since {\sc HoneyTech} is not publicly available. Notice that the time axis is logarithmic.

%\begin{figure}[h!]
%\begin{center}
%\includegraphics[width=2.0in]{fig/perfcompb.pdf}
%\end{center}
%\caption{Performance for TNC.}
%\label{fig:perf}
%\end{figure}

Because of the different hardware used, only a qualitative comparison can be made between our implementations and {\sc HoneyTech}:
going from 1 to 6 obstacles (as the case study in~\cite{honeytech}),
the run time of {\sc HoneyTech} shows an exponential behavior,
while our best implementation exhibits an approximately linear growth, as shown in
Figure~\ref{fig:perf}, where the performance of PHAVer+ is plotted up to 9 obstacles.

\paragraph{Water Tank Control.}
Consider the system depicted in Figure~\ref{fig:TanksDescr},
where two tanks --- A and B --- are linked by a one-directional valve \textit{mid} (from A to B). There are two additional valves: the valve \textit{in} to fill A and the valve \textit{out} to drain B. The two tanks are open-air: the level of the water inside also depends on the potential rain and evaporation.
It is possible to change the state of one valve only after one second since the last valve operation.

The corresponding hybrid automaton has eight locations, one for each combination of the state (open/closed) of the three valves, and three variables: $x$ and $y$ for the water level in the tanks, and $t$ as
the clock that enforces a one-time-unit wait between consecutive discrete transitions.
Since the tanks are in the same geographic location, rain and evaporation are assumed to have the same
rate in both tanks, thus leading to a proper LHA that is not rectangular~\cite{hybridgames99}.

%\begin{figure}[ht]
%\begin{center}
%\includegraphics[width=2.8in]{fig/TanksDescr.pdf}
%\end{center}
%\caption{Schema of Tanks and run time (in sec.) of three algorithms.}
%\label{fig:TanksDescr}
%\end{figure}

We set the \textit{in} and \textit{mid} flow rate to $1$, the \textit{out} flow rate to $3$, the maximum evaporation rate to $0.5$ and maximum rain rate to $1$, and solve the synthesis problem for the safety specification requiring the water levels to be between $0$ and $8$.
Figure~\ref{fig:TanksTime} shows the run time of the three versions of the algorithm on WTC.

\subsection{Micro Analysis} \label{sec:micro}
In this subsection we show the behavior of individual calls to
$\staymust_l{(Z,V)}$, implemented in the three different ways
described in Section~\ref{sec:sorm}. The evaluation of the efficiency
of the three versions is carried out based on the number of
comparisons that the three algorithms perform in order to identify the
boundaries between polyhedra in $Z$ and polyhedra in $\initentry$,
with respect to the size of the input.
%, i.e. the product between the cardinality of $Z$ and the cardinality
%of $\overline{Z}$.
We choose to highlight the number of computed boundaries
because the idea that led us to the realization of the final version
of the algorithm is precisely to avoid unnecessary
adjacency checks. % (hence the idea of introducing the adjacency relations).

\begin{figure}[h]
\begin{center}
\includegraphics[width=16.0cm]{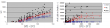}
\end{center}
\caption{Run time (in sec.) and number of boundary checks of the three algorithms for $\staymust$ w.r.t.\ the size of the input.}
\label{fig:micro}
\end{figure}

Figure~\ref{fig:micro}
shows the run time and the number of boundary computations made by the three
approaches. As expected, the number of calls made by
the basic algorithm is higher than those made by the adjacency
approach, which in turn is higher then those made by the local
adjacency algorithm. This is reflected in the execution times of the three procedures.
One also notices a certain instability in the case of the basic
algorithm,
due to the fact that in some instances of the problem, even with small
inputs, the algorithm can cut an individual polyhedron in many parts:
this dramatically increases the size of the sets $Z$ and
$\bar{Z}$ in the next steps and consequently the number of
comparisons required. This instability is held much more
under control with the introduction of the adjacency relations.
%, potentially
%preventing unnecessary checks. \\
Note that in the
local version the number of comparisons required is much lower: we
can easily explain this fact, recalling that $\initentry$ in
the adjacency version returns the whole $\bar{Z}$, 
forcing Algorithm~\ref{algo:updext} 
to perform $|\bar{Z}|$ iterations of its ``foreach'' loop.

\begin{figure}[h]
\begin{center}
\includegraphics[width=9.0cm]{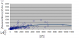}
\end{center}
\caption{Size of $\initentry$ in the \emph{Adj} and the \emph{Local}
algorithms.}
\label{fig:PotEntry}
\end{figure}

Figure~\ref{fig:PotEntry} shows, for the same inputs, the
relationship between the size of $\initentry$ in the basic and
in the adjacency versions (i.e., $\bar{Z}$) and in the local version:
the ratio is 1 to 10, which reduces
drastically the number of checks, and consequently the overall run time.

%\begin{figure}[ht]
%\begin{center}
%\includegraphics[width=9.0cm]{fig/sormtime.pdf}
%\end{center}
%\caption{Run time (in sec.) of the \emph{Basic}, the \emph{Adj} and the \emph{Local} algorithms.}
%\label{fig:sormtime}
%\end{figure}

%\begin{figure}[ht]
%\begin{center}
%\includegraphics[width=9.0cm]{fig/All1.pdf}
%\end{center}
%\caption{Number of boundary check w.r.t.\ input dimension.}
%\label{fig:All1}
%\end{figure}

%\begin{figure}[ht]
%\begin{center}
%\includegraphics[width=4.0in]{fig/All1.pdf}
%\end{center}
%\caption{Number of boundary check w.r.t.\ input dimension.}
%\label{fig:All1}
%\end{figure}

%\begin{figure}[ht]
%\begin{center}
%\includegraphics[width=4.0in]{fig/All2.pdf}
%\end{center}
%\caption{Number of boundary check w.r.t.\ input dimension.}
%\label{fig:All2}
%\end{figure}

%\setlength{\itemsep}{2cm}

% Trick: in order to compress bibliography,
% add this to the bbl file:
% \setlength{\parskip}{-5pt}

%\bibliographystyle{eptcs}
%\bibliography{../hybrid}

\end{document}